\documentclass[aip,jcp,amsmath,amssymb,reprint]{revtex4-1}
\usepackage{graphicx}
\usepackage{dcolumn}
\usepackage{bm}
\usepackage[utf8]{inputenc}
\usepackage[T1]{fontenc}
\usepackage{mathptmx}
\usepackage{etoolbox}
\usepackage{xcolor}

 \usepackage{multirow}

\usepackage{ulem}

\begin{document}

\title{
Time-resolved solvation of alkali ions in superfluid  helium nanodroplets: 
Theoretical simulation of a pump-probe study
} 

\author{Ernesto Garc\'{\i}a-Alfonso}
\affiliation{Laboratoire Collisions, Agr\'egats, R\'eactivit\'e  (LCAR), Universit\'e de Toulouse, CNRS, 31062, Toulouse, France}
  \affiliation{New affiliation: Institut des Sciences Mol\'eculaires d'Orsay (ISMO), CNRS and Universit\'e Paris-Saclay, 91400 Orsay, France}

\author{Manuel Barranco$^{\dag}$}
\affiliation{Departament FQA, Facultat de F\'{\i}sica, Universitat de Barcelona, Av.\ Diagonal 645, 08028 Barcelona, Spain.}
\affiliation{Institute of Nanoscience and Nanotechnology (IN2UB), Universitat de Barcelona, Barcelona, Spain.}
\affiliation{$^\dag$ deceased}

\author{Mart\'\i{} Pi}
\affiliation{Departament FQA, Facultat de F\'{\i}sica, Universitat de Barcelona, Av.\ Diagonal 645, 08028 Barcelona, Spain.}
\affiliation{Institute of Nanoscience and Nanotechnology (IN2UB), Universitat de Barcelona, Barcelona, Spain.}

\author{Nadine Halberstadt}
\email{Nadine.Halberstadt@irsamc.ups-tlse.fr}
\affiliation{Laboratoire Collisions, Agr\'egats, R\'eactivit\'e  (LCAR), Universit\'e de Toulouse, CNRS, 31062, Toulouse, France}

\begin{abstract}
The solvation process of an alkali ion (Na$^+$, K$^+$, Rb$^+$, Cs$^+$) inside a superfluid $^4$He$_{2000}$ nanodroplet is investigated theoretically using liquid $^4$He time-dependent density functional theory at zero temperature. 
We simulate both steps of the pump-probe experiment conducted on Na$^+$ [Albrechtsen \textit{et al.}, Nature \textbf{623}, 319 (2023)], where the alkali atom residing at the droplet surface is ionized by the pump pulse and its solvation is probed by ionizing a central xenon atom and detecting the expulsed Na$^+$He$_n$ ions. 
Our results confirm the  Poissonian model for the binding of the first five He atoms for the lighter Na$^+$ and K$^+$ alkalis, with a rate in good agreement with the more recent experimental results on Na$^+$ [Albrechtsen \textit{et al.}, J. Chem. Phys. \textbf{162}, 174309 (2025)]. 
For the probe step we show that the ion takes several picoseconds to get out of the droplet. 
During this rather long time, the solvation structure around it is very hot and far from equilibrium, and it can gain or lose more He atoms. 
Surprisingly,  analysing  the Na$^+$ solvation structure energy reveals that it is not stable by itself during the first few picoseconds of the solvation process. 
After that, energy relaxation follows a Newton behavior, as found experimentally, but with a longer time delay, $5.0\leq t_0\leq 6.5$~ps \textit{vs.} $0.23\pm0.06$~ps,  and characteristic decay time, $7.3\le\tau\le 16.5$~ps \textit{vs.} $2.6\pm 0.4$~ps. 
We conclude that the first instants of the solvation process are highly turbulent and that the solvation structure is stabilized only by the surrounding helium ``solvent'.

\end{abstract}

\maketitle

\section{INTRODUCTION}

Ions have been extensively used to probe superfluid liquid helium properties by determining their mobility.\cite{Tabbert1997,Foerste1997,Rossi2004,Fiedler2012} 
Due to the powerful attraction between helium and  positive charges, positive ions locally perturb the superfluid so strongly that they create a highly structured solvation structure around them, composed of very inhomogeneous, high-density helium layers.
This structure is  commonly named \textit{snowball}\cite{Atkins1959} because of its solid-like character.

Helium nanodroplets are nano-sized droplets produced in nozzle beams.
They have been shown to exhibit superfluid properties at their very low temperature 0.37~K (for a recent review, see Ref.~\onlinecite{Toennies2022}).
The solvation of ions in these finite-sized objects has been the subject of intense research activity, which has been recently reviewed by Gonz\'alez-Lezana \textit{et al.}\cite{GonzalezLezana2020}.
In particular, pure droplets have been found to be able to accommodate a surprisingly high density of charges close to their surface upon multiple ionization,\cite{Laimer2019,GarciaAlfonso2024b} with Coulomb explosion producing only small ions when the charge reached a critical value.\cite{Laimer2019}

Recently, helium nanodroplets have also been used as a convenient model solvent to study the dynamics of solvation on a microscopic scale.\cite{Albrechtsen2023,Albrechtsen2025}
Solvation is an ubiquitous process which is rather well understood on a macroscopic scale.
However, it has eluded a microscopic description because of the difficulty of observing directly as a function of time how solvent molecules approach a dissolving molecule and form a solvation complex.
An innovative experiment by Albrechtsen \textit{et al.}\cite{Albrechtsen2023,Albrechtsen2025} has shed some light on the dynamics of this process.
Using nanometer-size droplets of liquid helium as a model solvent, it aimed at monitoring the number of He atoms attaching to an ion, namely Na$^+$ or another alkali  ion (Ak$^+$), as a function of time.
Alkali atoms are known to reside in a dimple at the droplet surface.\cite{Stienkemeier1996,Buenermann2007,Pifrader2010}.
Upon ionization, they were shown to remain attached to the droplet\cite{Theisen2010,Loginov2011a,Theisen2011a,Theisen2011b}
The experimental confirmation for the complete solvation of a cation has come from photoionizing Ba at the surface of a helium droplet and comparing  the $6{}^2\text{P}\gets 6{}^2\text{S}$ absorption spectrum of Ba$^+$ from the one in bulk  He~II.\cite{Zhang2012}

The principle of the experiment by Albrechtsen \textit{et al.}\cite{Albrechtsen2023,Albrechtsen2025} is based on a femtosecond pump-probe set-up, the natural time scale for solvation.
Each helium nanodroplet is doped by one Ak atom and one Xe atom.
Due to a stronger helium-atom interaction, the xenon atom sits at or near the center of the droplet,\cite{Ancilotto1995b,Poms2012,Coppens2016} while as mentioned earlier, the Ak atom sits in a dimple at its surface.
A femtosecond laser pulse (``pump'' pulse) ionizes the Ak atom, which marks the beginning of the Ak$^+$ ion solvation.
Ak$^+$-He interaction being strongly attractive, the newly formed ion moves towards the center of the droplet and more and more He atoms bind to it.
After a variable time delay $\Delta t$, a second femtosecond laser pulse (``probe'' pulse) ionizes the Xe atom at the droplet center.
This triggers Coulomb repulsion between the two ions, and the Ak$^+$ ion is ejected, carrying  its solvation shell along.
Detecting the Ak$^+$He$_n$ complexes as a function of the time delay $\Delta t$ provides a way to follow the time evolution of the build-up of the solvation shell.
The binding dynamics of the first five helium atoms was found to be well described by a Poissonian process with a binding rate of 2.0 atoms per picosecond.
This rate was consistent with time-dependent helium-density functional theory ($^4$He-TDDFT) calculations that accompanied the experimental results.\cite{Albrechtsen2023}
A more comprehensive and detailed account\cite{Albrechtsen2025} of the experimental results concluded that this binding rate depended on the average droplet size distribution $\langle N_D \rangle$: 
$1.65\pm 0.09$~atom/ps for $\langle N_D \rangle=3600$; 
 $1.84\pm 0.09$~atom/ps  for $\langle N_D \rangle=5200$; 
  $2.04\pm 0.13$~atom/ps  for $\langle N_D \rangle=9000$.

Albrechtsen \textit{et al.}\cite{Albrechtsen2025} have also used a model to determine the energy dissipated during the solvation process from the local region around the ion to the rest of the droplet.
By fitting the time-dependent size distributions of the detected Na$^+$He$_n$ ions,
they found that the mean dissipated energy could be fitted with Newton's law of cooling for the first 5~ps of the solvation process. This law assumes that the rate of energy transfer from the hot solvation structure to the surrounding helium droplet  is proportional to its internal energy.

From the point of view of theory, several studies were conducted in order to visualize the solvation process, as well as to confirm some of the hypotheses on which the experimental results were based.
A first publication\cite{Leal2014} made use of the helium time-dependent density functional theory ($^4$He-TDDFT) approach to investigate the  dynamics following the photoionization of neutral Rb and Cs atoms residing in a dimple at the surface of a superfluid $^4$He$_{1000}$ nanodroplet. 
The calculations revealed that structured high density helium solvation layers formed around  the Rb$^+$ or the  Cs$^+$ cation on a picosecond time scale, building the so-called snowballs. 

The second one, which was included in a common publication with experiment,\cite{Albrechtsen2023} was also based on the $^4$He-TDDFT  approach.
It was describing the time evolution of the helium density around the solvating Na$^+$ ion, and concluded that indeed the time required for $n$ atoms to move inside the first or the second solvation shell behaved linearly with $n$, at least up to $n=5$, therefore validating the Poissonian model for the binding of the first He atoms.
The experimental times were found to lie in between the results of the simulations for the first and the second solvation shells.

These results were further confirmed in a more complete $^4$He-TDDFT theoretical study\cite{GarciaAlfonso2024} including the solvation of Li$^+$, Na$^+$, K$^+$, Rb$^+$, and Cs$^+$ ions.
The linear behavior for the time taken by $n$ He atoms to get into the first solvation shell up to $n=5$ was confirmed for all alkali ions. 
In particular, the rate $A$ obtained from the fit   $n(t)=At$ of the simulations was found to be between 0.74 and 0.79~atom~ps$^{-1}$ for Na$^+$.
In addition, an interesting kinetic effect was observed for the two lighter alkalis, Li$^+$ and Na$^+$.
The number of He atoms in the first solvation shell was found to converge after 20 ps (Li$^+$) or 25-30~ps (Na$^+$) to a smaller number than at equilibrium:
(9 instead of 12 for Li$^+$, 12 instead of 14 for Na$^+$), even though the simulations were conducted for longer times.
This was tentatively attributed to the rigid character of the solvation structure around these ions, and in particular to the clear separation between the first and the second solvation shells.
The continuous exchange of helium density between these two shells that occurred in the case of the heavier alkalis could no longer take place when the shells were separated, which occurred after  $\sim$20~ps for Li$^+$ and $\sim$25-30~ps for Na$^+$.

A recent study by Calvo\cite{Calvo2024,Calvo2025} using a completely atomistic approach, the phenomenological path-integral Monte Carlo (PIMC) method, on Na$^+$ and K$^+$ at a temperature of 1~K or 2~K and without boson exchange symmetry.
This computational work confirms a fast initial capture of He atoms by the Ak$^+$ ion, although the capture is delayed by a few picoseconds in the case of K$^+$.

The $^4$He-TDDFT simulations described above were focusing on the solvation process, which is the first step (pump stage) of the experiment by Albrechtsen \textit{et al.}\cite{Albrechtsen2023}\ 
However, the number of He atoms attached to the ions are detected at the pump stage, \textit{i.e.,} once the Xe atom at the droplet center has been ionized after a pump-probe delay of $\Delta t$, thus triggering the ejection of the Ak$^+$ ion and its surrounding helium atoms, and once this structure has reached the detecting region.
Therefore, the Ak$^+$He$_n$ complex detected could differ from the Ak$^+$He$_N$ complex that was ejected from the droplet : $n\neq N$.
Several processes can in principle play a role:
(i) the Ak$^+$ ion has been travelling towards the droplet center during the pump stage, it takes some time for it to slow down, turn around and get out of the droplet again; During that time, its solvation structure can bind or lose He atoms and its internal energy can increase by collisions with the rest of the droplet or decrease by dissociating He atoms; 
(ii) on its way towards the detector, the solvation structure being internally warm can relax by dissociating helium atoms

In this work we model the two stages of the experiment for a realistic droplet size of 2000 He atoms.
The solvation (pump) stage repeats the earlier study, albeit with a Xe atom at the droplet center and with a finer simulation grid for a higher accuracy.
The dynamics following the probe pulse is also simulated.
Due to computational costs, we limit the probe study to the case where the solvation structure has reached 5 He atoms.
This is the largest size for which the Thomson model was validated:  above that size the ion yield signals were influenced by the dissociation of higher complexes.
Section~\ref{sec:theory} describes the main elements of the $^4$He-TDDFT approach in the context of the pump-probe simulations; 
Section~\ref{sec:res} presents the results of the solvation (pump) and Coulomb-driven ion separation (probe) step; 
In Section~\ref{sec:energy} the energetics of the solvation structure is analyzed during the solvation step for Na$^+$, for which there is an experimental counterpart\cite{Albrechtsen2025};
Finally, Section~\ref{sec:conclusion} summarizes the results and give some concluding remarks.
Note that movies illustrating the real time simulation of the pump-probe process are provided in the Supplementary Material.
This multimedia material provides  physical insight that would be difficult to describe in detail in the main text.

\section{Theoretical method:  $^4$He-(TD)DFT }\label{sec:theory}

We use the density functional theory approach applied to helium density, denoted as  $^4$He-(TD)DFT,
to simulate  the Ak atom ionization (pump step) and the following Xe atom ionization (probe step) in a $^4$He-droplet.
$^4$He-(TD)DFT  is a compromise between accuracy and feasiblility, which has proven to be accurate and powerful in a number of cases.\cite{Barranco2006,Ancilotto2017,Coppens2018,Rendler2021,Trejo2024}
The Orsay-Trento (OT) functional\cite{Dalfovo1995} has been phenomenologically determined to reproduce superfluid behavior, namely, boson exchange symmetry, exchange and correlation effects, superfluid helium elementary excitation curve...
The method itself being well documented,\cite{Dalfovo1995,Barranco2006,Ancilotto2017,Barranco2017}  we only give here details specific to the application at hand.

All the simulations presented in this work involve a droplet of 2000 atoms: this size was chosen as a compromise between feasibility and relevance to experimental sizes. 
The alkali atoms at the surface, which are ionized in the pump step, are Na, K, Rb, and Cs.
The central atom which is ionized in the probe step is Xe, as in experiment.\cite{Albrechtsen2023,Albrechtsen2025}

\subsection{Equilibrium Ak@(Xe@He$_N$) configuration (Statics)}

Within the $^4$He-DFT approach at zero temperature, the energy of a $N_D$-atom  helium droplet $^4$He$_N$ 
doped with an Ak  and a Xe atoms, both treated classically (\textit{i.e.}, represented by an external field), is written as a functional of the $^4$He atom 
density $\rho({\mathbf r})$ as:
\begin{eqnarray}
E[\Psi] &=& \int d \mathbf{r} \, \frac{\hbar^2}{2m_{\rm He}}|\nabla \Psi|^2 +  \int d \mathbf{r} \, {\cal E}_c(\rho)   
\nonumber\\
&+& \sum_{\text{At}\equiv\text{Ak, Xe}}\int d \mathbf{r} \,\, V_{\rm He-At}(|\mathbf{r}-\mathbf{r}_{\rm At}|) \, \rho(\mathbf{r}) 
\label{eq:E-HeDFT}\\
&+& V_\text{Xe-Ak}(|\mathbf{r}_{\text{Xe}}-\mathbf{r}_{\rm Ak}|),
\nonumber
\end{eqnarray}
where the first term is the kinetic energy of the superfluid, 
$m_{\rm He}$ is the mass of the $^4$He atom, and
$\Psi({\mathbf r})$ is the effective wave function (or order parameter) of the superfluid, normalized such that 
$\rho({\mathbf r})=|\Psi({\mathbf r})|^2$ with $\int d{\bf r}|\Psi({\bf r})|^2 =N_D$. 
The  ${\cal E}_c(\rho)$ functional includes a He-He interaction term within the Hartree approximation and additional terms describing non-local correlation effects.
Since the same functional has to be used in the dynamics, which involves ions with a strong attractive interaction with helium,
the version of $\mathcal{E}_c$ used in this work is a modification\cite{Ancilotto2005a} of the Orsay-Trento functional\cite{Dalfovo1995} which makes it stable even in the presence of very attractive dopants. 

All the potentials in Eq.~(\ref{eq:E-HeDFT} are approximated by sums of atom-atom interactions, taken from the literature.
Their characteristics and references are collected in the left part of Table~\ref{Tab:pots}.
Most of them (except for Xe)  are illustrated in Fig.~2 of Garc\'\i a-Alfonso et al.\cite{GarciaAlfonso2024}
The Xe-Ak potentials, last line in Eq.~(\ref{eq:E-HeDFT}), were not used since the Xe atom is expected to sit at the droplet center and the Ak atom at its surface: at that distance (sharp density radius of $^4$He$_{2000}$ = 28.0~\AA), the Xe-Ak interaction is negligible.
All the He-Ak potentials were taken from Patil\cite{Patil1991} in order to keep the same accuracy level for all alkalis.

 \begin{table}
\begin{center}
\begin{tabular}{ccccc | ccccc}
\hline
\hline
 & $D_e$ (K)  & $R_e$ (\AA)   &  $R_0$(\AA) & Ref.  &  & $D_e$ (K)  & $R_e$ (\AA)   &  $R_0$(\AA) & Ref. \\
\hline
 Li       & 1.93 & 6.19 & 5.49 & \onlinecite{Patil1991} & Li$^+$       & 852.1 & 1.92 & 1.59 & \onlinecite{Koutselos1990}  \\ 
 Na     & 1.73 & 6.42 & 5.69 & \onlinecite{Patil1991} &  Na$^+$     & 410.3 & 2.41 & 2.04 & \onlinecite{Koutselos1990} \\ 
 K       & 1.40 & 7.18 & 6.37 & \onlinecite{Patil1991}  &  K$^+$       & 236.5 & 2.90 & 2.50 & \onlinecite{Koutselos1990} \\ 
 Rb     & 1.41 & 7.33 & 6.51 & \onlinecite{Patil1991}  &  Rb$^+$     & 204.1 & 3.10 & 2.68 & \onlinecite{Koutselos1990}\\ 
 Cs     & 1.21 & 7.73 & 6.86  & \onlinecite{Patil1991} &  Cs$^+$     & 168.9 & 3.37 & 2.95 & \onlinecite{Koutselos1990} \\ 
 Xe     & 28.1 & 3.98 & 3.55  & \onlinecite{Sheng2020}  &  Xe$^+$     & 206.5 & 3.24 & 2.80  &  \onlinecite{Viehland2009} \\
\hline
\hline
\end{tabular}
\end{center}
\caption{
Parameters of the atom-helium and ion-helium interaction potentials used in this work:
dissociation energy $D_e$, equilibrium distance $R_e$, and distance $R_0$ at which the potential is zero.
\label{Tab:pots}
}
\end{table}
 
The equilibrium configuration of the doubly-doped droplet is obtained by solving the Euler-Lagrange equation arising 
from functional variation of Eq. (\ref{eq:E-HeDFT}):
\begin{eqnarray}
{\cal H}[\rho] \,\Psi  
 &=& \mu \Psi,
 \label{eq:HeDFT-eq} \\
\text{where} \,\,\,{\cal H}[\rho]  &=& -\frac{\hbar^2}{2m_{\rm He}} \nabla^2 + \frac{\delta {\cal E}_c}{\delta \rho} + 
V_{\rm He-Ak}(|\mathbf{r}-\mathbf{r}_{\rm Ak}|)
\nonumber\\
&&+ V_{\rm He-Xe}(|\mathbf{r}-\mathbf{r}_{\rm Xe}|) + V_{\rm Xe-Ak}(|\mathbf{r}_{\rm Xe}-\mathbf{r}_{\rm Ak}|).
\label{eq:H-HeDFT}
\end{eqnarray}
In Eq.~(\ref{eq:HeDFT-eq}), $\mu$ is the $^4$He chemical potential corresponding to the number of He atoms in the droplet 
($N=2000$ in this work, corresponding to a droplet of $R= 28$ \AA{} radius\cite{Barranco2006}), 
and $\cal{H}$ is the DFT Hamiltonian detailed in Eq.~(\ref{eq:H-HeDFT}). 
The Xe atom is placed at the droplet center and the Ak atom at its equilibrium position in a $^4$He$_{2000}$ droplet.
The Euler-Lagrange equation is then solved by a relaxation (imaginary time) method using the $^4$He-DFT BCN-TLS 
computing package,\cite{Pi2024} see  Refs.~\onlinecite{Ancilotto2017} and \onlinecite{Barranco2017} 
and references therein for additional details.
$\Psi(\mathbf{r},t)$ is defined at the nodes of a 3-dimension Cartesian grid with 0.2~\AA{} space step.
The whole simulation box is 108.6$\times$108.6$\times$127.8~\AA\ ($544 \times 544 \times 640$ points), with the drop center initially located at $(0,0,-6)$~\AA.
Periodic boundary conditions are imposed in order to efficiently calculate the convolutions involved in the  mean field ${\cal H}[\rho]$ using Fast Fourier  Transform\cite{Frigo2005}. 
The differential operators  in ${\cal H}[\rho]$ are approximated by 13-point formulas.

\subsection{Dynamics}\label{ssec:meth-dyn}

\begin{figure}[t]
\includegraphics[width=1.0\linewidth,angle=0,clip=true]{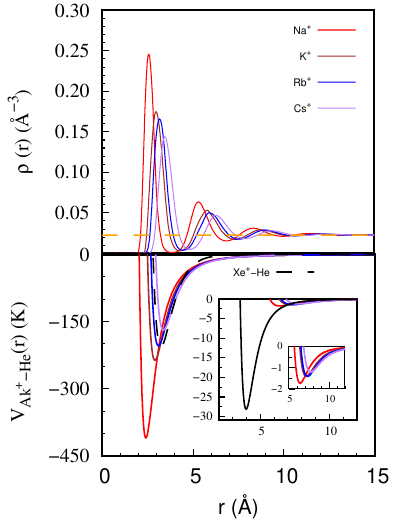}
\caption{\label{fig:Akp-He-pots-1D-dens}
Bottom: Ak$^+$-He (see color key in the upper plot) and Xe$^+$-He (dashed black line) interaction potentials;
The insert shows the corresponding neutral interaction potentials.
Top: 1-dimension helium density profile for the equilibrium configuration of Ak$^+$@$^4$He$_{2000}$,
helping to visualize the first and second solvation shells of the ions.
The density of the pure He$_{2000}$ droplet, equal to the bulk superfluid helium density at 0~K ($\rho_0=0.0218$~\AA$^{-3}$) in this interval of distance from the central ion, is represented as a yellow dashed line.}
\end{figure}

In the experiment by Albrechtsen \textit{et al.},\cite{Albrechtsen2023,Albrechtsen2025} the dynamics proceeds in two stages. 
The pump stage is triggered by ionizing the alkali atom on the droplet surface with a 100-fs laser pulse.
This is reproduced in our simulations by  assuming sudden ionization:
in practice, the Ak atom is replaced by its ion Ak$^+$ in  Eq.~(\ref{eq:E-HeDFT}) expressing the total energy of the system.
The equations of motion are then obtained by minimizing the action,\cite{Ancilotto2017,GarciaAlfonso2024}\ which gives three coupled equations describing the time evolution of the helium pseudo-wavefunction and of the alkali ion and the xenon atom positions:
\begin{eqnarray}
i \hbar\frac{\partial}{\partial t} \Psi &=&
\left\{
  -\frac{\hbar^2}{2m_{\rm He}}\nabla^2 +
  \frac{\delta {\cal E}_c}{\delta \rho(\mathbf{r})} 
  + V_{\text{He-Ak}^+}(|\mathbf{r}- \mathbf{r}_{\rm Ak^+}|) 
  \right.\nonumber\\
&&\left. + V_{\text{He-Xe}}(|\mathbf{r}- \mathbf{r}_{\text{Xe}}|) \right\} \Psi;
\label{eq:HeTDDFT-pump1}
\\
m_{\rm Ak^+}\,\ddot{\mathbf{r}}_{\rm Ak^+} &=& 
  -  
  \int d \mathbf{r} \,  V_{\text{He-Ak}^+}(|\mathbf{r}- \mathbf{r}_{\text{Ak}^+}|)  \nabla \rho(\mathbf{r}) 
  \nonumber\\
 && -\nabla_{\mathbf{r}_{\text{Ak}^+}}  V_{\text{Xe-Ak}^+}(|\mathbf{r}_{\text{Xe}}-\mathbf{r}_{\text{Ak}^+}|);
\label{eq:HeTDDFT-pump2}
\\
m_{\text{Xe}}\,\ddot{\mathbf{r}}_{\text{Xe}} &=& 
  -  
  \int d \mathbf{r} \,  V_{\text{He-Xe}}(|\mathbf{r}- \mathbf{r}_{\text{Xe}}|)  \nabla \rho(\mathbf{r}) 
  \nonumber\\
 && -\nabla_{\mathbf{r}_{\text{Xe}}}  V_{\text{Xe-Ak}^+}(|\mathbf{r}_{\text{Xe}}-\mathbf{r}_{\text{Ak}^+}|),
\label{eq:HeTDDFT-pump3}
\end{eqnarray}
where  the time dependence of the variables  has been omitted for clarity. 
The initial conditions are the ones obtained from the statics simulation, i.e., the equilibrium configuration and helium density of the Ak@[Xe@ $^4$He$_{2000}$] droplet.
All the ion-helium potentials are approximated by a sum of atom-atom interactions, taken from the literature.
The characteristics and references for Ak$^+$-He interaction potentials are collected in the right-hand part of Table~\ref{Tab:pots}.
They are illustrated in Fig.~\ref{fig:Akp-He-pots-1D-dens}, together with the equilibrium density profile for Ak$^+$ at equilibrium in a 2000-atom droplet.
All the He-Ak$^+$  potentials were taken from Koutselos\cite{Koutselos1990} in order to keep the same accuracy level for all alkalis.
Note that they include the correct $1/R^4$ behavior corresponding to charge-induced dipole interaction at long range.

 
The probe stage in the experiment is triggered by ionizing the xenon atom after a given time delay $\Delta t$ from the pump.
This is simulated by assuming sudden ionization of the Xe atom after running the probe simulation for $\Delta t$,
\textit{i.e.}, by replacing  $V_{\text{He-Xe}}$ with $V_{\text{He-Xe}^+}$ (referenced in Table~\ref{Tab:pots}) in Eq.~(\ref{eq:HeTDDFT-pump1}), 
   $V_{\text{Xe-Ak}^+}$ with $V_{\text{Xe}^+\text{-Ak}^+}$ (taken as Coulomb charge-charge repulsion),
 and $\mathbf{r}_{\text{Xe}}$  with $\mathbf{r}_{\text{Xe}^+}$
 in Eqs.~(\ref{eq:HeTDDFT-pump1}-\ref{eq:HeTDDFT-pump3}).
 
 The same simulation box  as for the statics is used for all the dynamics.
 Unphysical effects could arise as  a consequence of using  periodic boundary conditions:  Helium density reaching the simulation box boundaries would re-enter from the opposite side and interfere with the droplet density.
 This is especially true here since the ion solvation process is highly energetic and dissipates a lot of energy (several thousand K, see Eq.(7) and Table I in Ref.~\onlinecite{GarciaAlfonso2024}).
 In order to avoid that, an absorption potential is added inside a buffer starting 2~\AA{} away from the box limits: it gradually drives the density to zero at the boundaries. 
 More details can be found in Refs.~\onlinecite{Ancilotto2017,Barranco2017}.

 
\begin{figure}[t]
\includegraphics[width=1.0\linewidth,angle=0,clip=true]{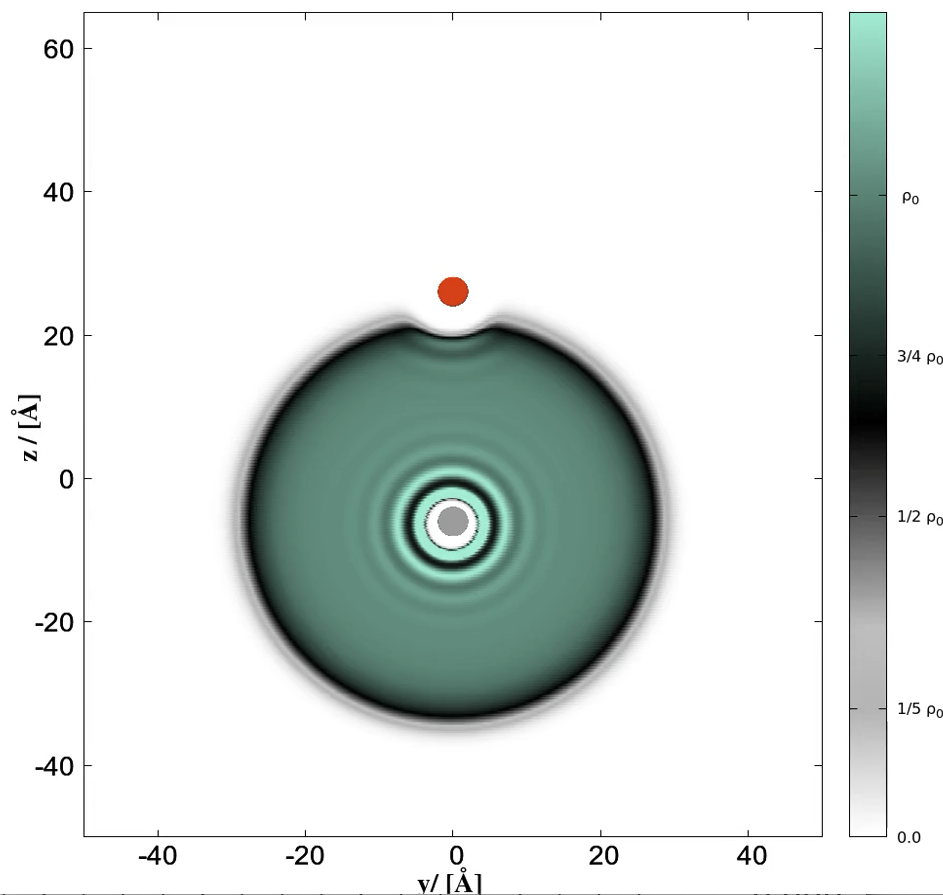}
\caption{\label{fig:Ak-Xe-at-He2000}
Equilibrium configuration of Rb@(Xe@$^4$He$_{2000}$) droplet (2-dimension cut).
The Rb-Xe distance is $32.1$~\AA.
The density scale on the right is given in units of $\rho_0$, the bulk superfluid helium density at zero temperature and pressure
($\rho_0=0.0218$~\AA$^{-3}$).
The sharp density radius of He$_{2000}$ is $28.0$~\AA.}
\end{figure}


 

\begin{figure*}[t]

\hspace*{-0.5cm}
\includegraphics[angle=270,width=1.05\linewidth,clip=true]{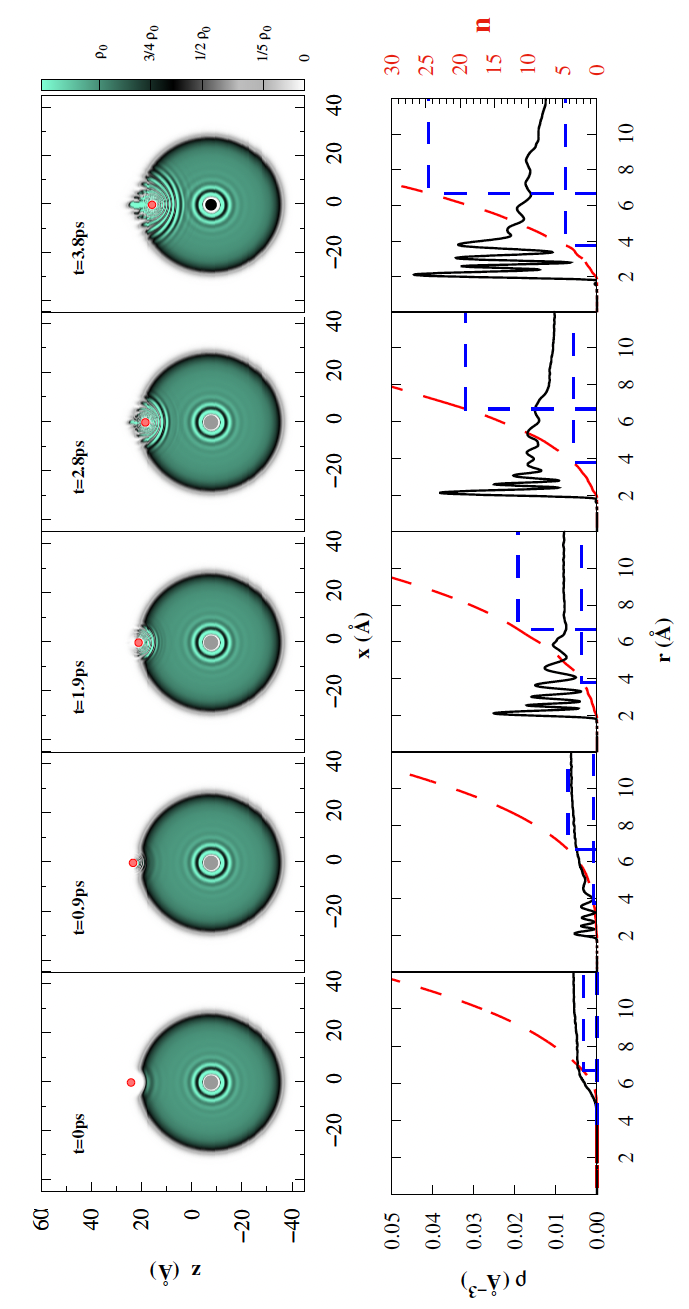}\\

\vspace{0cm}
\caption{\label{fig:pump-Nap-snapshots}
Snapshots every $\equiv0.95$~ps taken during the pump step of Na$^+$@(Xe$^+$@$^4$He$_{2000}$) droplet
(the last snapshot at $t=3.8$~ps corresponds to the probe time, at which the Xe atom is ionized: it is then represented by a smaller, blue dot, reflecting the shorter He-Xe$^+$ equilibrium distance).
Top: 2-dimension cuts ; 
Bottom: spherically integrated density as a function of the distance to the Na$^+$ ion.
The radius of the first  (3.76~\AA) and second  (6.68\AA) solvation shells are indicated as a vertical dotted blue line,
and the corresponding number of included He atoms are reported as a horizontal dotted blue line referred to the right hand vertical axis.
The complete movie is included as supplementary material.
}

\end{figure*}

\begin{figure*}[t]

\hspace*{-0.5cm}
\includegraphics[angle=270,width=1.05\linewidth,clip=true]{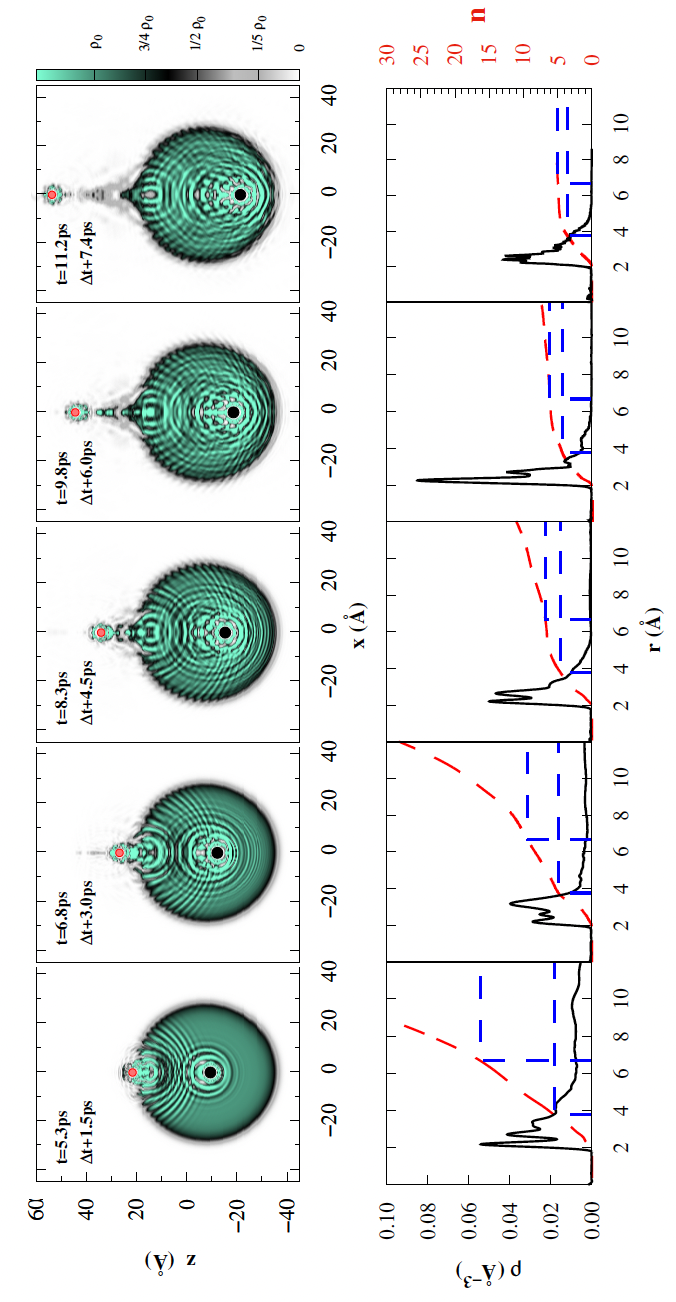}\\

\vspace{0cm}
\caption{\label{fig:probe-Nap-snapshots}
Snapshots every 1.5 ps taken during the probe step of Na$^+$@(Xe$^+$@$^4$He$_{2000}$) droplet.
Probe started at $t=3.8$~ps by ionization of the central Xe atom, triggering Coulomb repulsion between the two ions.
Top: 2-dimension cuts ; 
Bottom: spherically integrated density as a function of the distance to the Na$^+$ ion.
The radius of the first  (3.76~\AA) and second  (6.68\AA) solvation shells are indicated as a vertical dotted blue line,
and the corresponding number of included He atoms are reported as a horizontal dotted blue line referred to the right hand vertical axis.
The complete movie is included as supplementary material.
}

\end{figure*}



  \section{Results}\label{sec:res}
 The equilibrium configuration obtained from the static simulation is illustrated in Fig.~\ref{fig:Ak-Xe-at-He2000} for Rb@(Xe@He$_{2000}$).
The main difference with the helium density presented in Fig.~1 of Garc\'\i a-Alfonso \textit{et al.}\cite{GarciaAlfonso2024} is the presence of the Xe atom  at the droplet center, inducing a layered solvation structure around it.
However, the bulk helium density is recovered beyond about three layers, and the equilibrium density around the Ak atom is not affected.

In two preceding publications,\cite{Albrechtsen2023,GarciaAlfonso2024}\ the early stages of the solvation process of the Ak$^+$ ions were studied in simulations that are equivalent to the ones presented here for the pump stage.
The Xe atom at the droplet center was omitted, as it was considered to be far enough from the Ak$^+$ ion for its role to be neglected.
The first solvation shell of each Ak$^+$ ion was defined by considering the helium density profile around Ak$^+$ in a 2000-atom drop at equilibrium.
It was taken as the density peak closest to the cation, clearly separated from the rest of the helium density (see Fig.~2 in Ref.~\onlinecite{GarciaAlfonso2024}).
The radius $r_1^e$ of the  first solvation shell was then defined as the distance from Ak$^+$ at which the helium density was minimum (equal or close to zero) after the first density peak.
The number $n^e_1$  of helium atoms in the first solvation shell at equilibrium was then obtained by integrating the helium density inside a sphere of radius $r^e_1$ around Ak$^+$.
In this work we also make use of the second solvation shell, defined as the helium density included in the second density peak: 
The outer radius $r^e_2$ of the second solvation shell is defined as the distance of the density minimum following the second density peak to the ion.
We then refer to the the first or second solvation \textit{structures}: the first one is equivalent to the first solvation shell, and the second one is the reunion of the first and second solvation shells.
For the sake of completeness, $r^e_1$, $r^e_2$, and $n^e_1$ are collected in Table~\ref{Tab:R1n1dt}.


 \begin{table}
\begin{center}
\setlength{\tabcolsep}{6pt} 
\begin{tabular}{cccccccc}
\hline
\hline
 & $r^e_1$ & $n_1^e$  & $r^e_2$ &   $A$          & $\Delta t$ & $d_0$ & $d_{\Delta t}$ \\
 &  (\AA) &                &  (\AA)  &  (ps$^{-1}$) &  (ps)        & (\AA)                                 &  (\AA)  \\
\hline
 Na$^+$ & 3.8  & 14 & 6.7 &  [1.3, 1.3] & 3.8 & 31.9 & 23.5 \\
 K$^+$   & 4.3  & 17 & 7.3 &  [1.0, 1.0] & 5.2 & 32.5 & 25.3 \\
 Rb$^+$ & 4.5  & 18 & 7.5 & [1.1, 0.5] & 8.2 & 32.1 & 25.6 \\
 Cs$^+$ &  4.8 & 21 & 7.9 & [0.7, 0.7] & 8.1 & 32.1 & 26.4 \\
\hline
\end{tabular}
\end{center}
\caption{
Radius $r^e_1$ of the first solvation shell of Ak$^+$@He$_{2000}$ and number $n^e_1$ of  helium atoms in this shell at equilibrium, from Garc\'\i a-Alfonso \textit{et al.}\cite{GarciaAlfonso2024} (see text);
Radius $r^e_2$ of the second solvation shell;
Slope $A$ of the $n_1(t)=At$ fit of $n_1(t)\leq 5$ (see text for explanation of the two values);
Time delay $\Delta t$ at which the probe is started, \textit{i.e.}, time at which $n_1(t)=5$;  
Initial value, $d_0$, and value at $\Delta t$, $d_{\Delta t}$, of the Ak-Xe distance, the Xe atom sitting initially at the helium center of mass.
\label{Tab:R1n1dt}
}
\end{table}
 

\subsection{Pump (solvation) step dynamics}\label{ssec:pump}

\subsubsection{Description of the dynamics}\label{sssec:pump-description}
Figure \ref{fig:pump-Nap-snapshots} illustrates the dynamics during the pump step, 
\textit{i.e.} following ionization of the alkali, here Na as an example.
The corresponding movies showing the time evolution of the solvation of Na$^+$ and of the other alkali ions are collected as Supplementary Material.
The top part of the figure shows snapshots of the ion position in  2-dimensional cuts through the helium density in a plane containing both the Ak$^+$ ion and the Xe atom. 
The lower part shows the corresponding snapshots of the  helium density integrated in a sphere of radius $r$ around the alkali ion as a function of $r$, as well as the number of helium atoms inside that sphere (dashed red curve, referred to the right vertical axis).
The values of $r_1^e$ and $r_2^e$, the radii  of the first and second solvation shells respectively, 
are also indicated as vertical, blue dashed lines for reference, and the numbers $n_1$ and $n_2$ of He atoms contained in the  first and second solvation structures are indicated as horizontal, blue dashed lines.

\begin{figure}[t]
\includegraphics[angle=270,width=1.0\linewidth,clip=true]{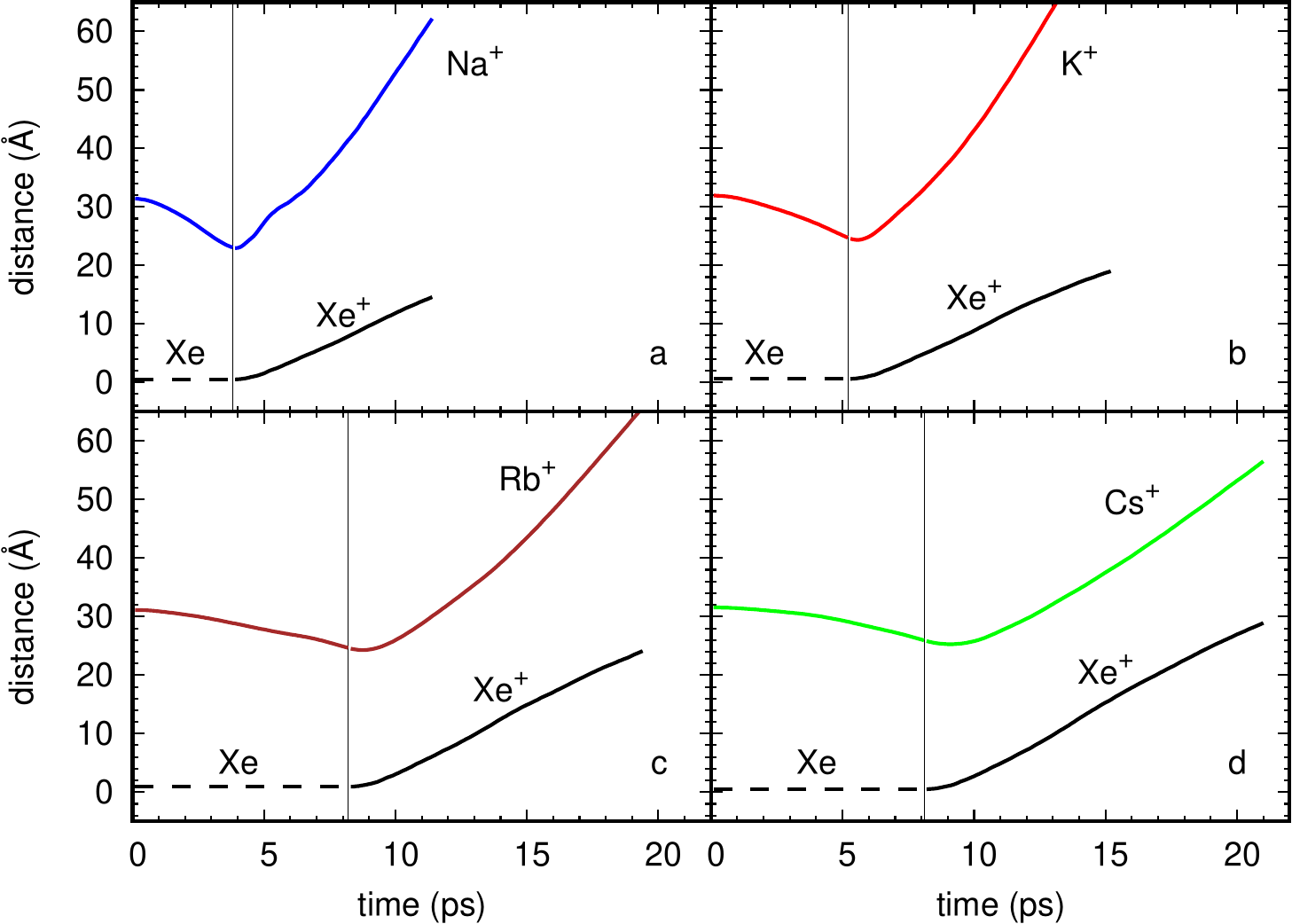}
\caption{\label{fig:Akp-Xep-HeCOMdist-t-all-Ak-pump-probe}
Time evolution of Ak$^+$ and Xe (pump, dashed line) or Xe$^+$ (probe, solid line) distance to the overall center of mass during the pump and the probe steps for each alkali.
The pump-probe delay $\Delta t$ is indicated as a black vertical line: it corresponds to $n_1=5$ in the pump step and triggers the probe (ionization of Xe).
}
\end{figure}

The behavior is qualitatively very similar to the one found by Garc\'\i a-Alfonso \textit{et al.}\cite{GarciaAlfonso2024}, where no xenon atom was present.
The Ak$^+$ ion tends to penetrate the droplet as more and more helium bind to it.
This can be quantified by plotting the distance of the Ak$^+$ ion to the center of mass of the system, Figs.~\ref{fig:Akp-Xep-HeCOMdist-t-all-Ak-pump-probe}.
During the solvation step (separated from the probe by a vertical line), all the ions penetrate inside the droplet in a monotonic way.
The lighter alkalis move faster, due to their lighter mass and stronger Ak$^+$-He attraction. 
The values of the Ak$^+$-Xe distances at the beginning ($d_0$) and the end [$d(\Delta t)$] of the pump stage are collected in Table~\ref{Tab:R1n1dt}.

The first solvation shell is clearly not stabilized when the number of He atoms reaches 5.
The spherically integrated density exhibits several oscillations instead of the single peak clearly separated from the rest of the density observed at equilibrium (see Fig.~\ref{fig:Akp-He-pots-1D-dens}).
This reflects a high internal energy in the Ak$^+$He$_{n_1}$ complex being formed.
As an additional sign of this high internal energy, the simulations in Garc\'\i a-Alfonso \textit{et al.}\cite{GarciaAlfonso2024} showed the evaporation of highly energetic He atoms during the first 20-30~ps of the solvation dynamics.
This high internal energy is a consequence of the sinking energy, \textit{i.e.}, the difference between the total energy at the ionization time and the one of Ak$^+$@$^4$He$_N$ at equilibrium:
4461.2~K, 3329.4~K, 3119.7~K and 2980.8~K for Na$^+$, K$^+$, Rb$^+$ and Cs$^+$ in a $^4$He$_{2000}$ droplet respectively\cite{GarciaAlfonso2024}.

\subsubsection{Binding rate}\label{sssec:binding-rate}

As a word of caution, here and wherever we refer to experimental results, we follow the notation used in Refs.~\onlinecite{Albrechtsen2023} and \onlinecite{Albrechtsen2025}:
$N$ refers to the ionic cluster size at probe time (\textit{i.e.,} when the Xe atom is ionized) and $n$ to the size of the one detected.
The two can in principle be different due to the time it takes for the ionic complex to separate from the droplet and reach the detector ($\sim$ms).

In order to interpret their time-dependent ion yields $Y_n(t)$, Albrechtsen \textit{et al.}\cite{Albrechtsen2023,Albrechtsen2025} considered the Poisson model in which the He atoms bind independently of each other and at a constant rate $A$.
In this model, the probability $P_N(t)$ for $N$ He atoms to have bound to the Ak$^+$ ion during the time interval $[ 0, t ] $ is given by
\begin{equation}\label{eq:Poisson}
P_N(t)=\frac{(At)^N e^{-At}}{N!},
\end{equation}
with the maximum of $P_N(t)$ occurring at $t=N/A$.

The experimentally measured ion yields $Y_n(\Delta t)$ have been used to check the validity of the Poisson model and to obtain the He atom binding rate, by comparing $Y_n(\Delta t)$  with $P_N(t)$ with $n=N$.
This was restricted to $n\le 5$ for two reasons.
(i) The He$_n$Ak$^+$ fragments with  $n>5$ were found to have a significant contribution from higher complexes which would dissociate during the flight time to the detection region due to their high internal energy.
This dissociation process is not taken into account by the Poisson model.
(ii) The binding energy of the first few He atoms was considered to be strong enough to prevent their dissociation before reaching the detector. As a consequence, $Y_n(t)$ could be directly compared to $P_N(t)$ for $N=n$.  

In their first publication,\cite{Albrechtsen2023}  Albrechtsen \textit{et al.} fitted the  position of the maxima of the $Y_n(t)$, $n\le 5$, peaks and concluded that they were following the linear behavior predicted by the Poisson model, with a rate of 2.0 atom/ps.
The validity of the Poisson model for $n\le5$ was further confirmed in their more recent and more complete experimental results,\cite{Albrechtsen2025} in which the complete $Y_n(t)$, $n\le 5$,  peaks were successfully fitted by the Poissonian form of Eq.~(\ref{eq:Poisson}) for three different droplet size distributions.
The binding rate was found to decrease with  decreasing droplet size:  
$2.04\pm 0.13$~atom/ps for $\langle N\rangle= 9000$,
$1.84\pm 0.09$~atom/ps for $\langle N\rangle= 5200$,
and $1.65\pm 0.09$~atom/ps for $\langle N\rangle= 3500$.

The $^4$He-TDDFT simulation included in Ref.~\onlinecite{Albrechtsen2023} confirmed the validity of the Poissonian model by obtaining a linear behavior for $n_i(t)$, the number of He atoms inside the first or second solvation shell, the experimental rate falling between the ones for $n_1$ and $n_2$ (see Fig.~3 in that paper).
In our previous work on Ak$^+$ solvation dynamics,\cite{GarciaAlfonso2024} Ak$^+ \equiv$ Li$^+$, Na$^+$, K$^+$, Rb$^+$ or Cs$^+$,
 the time taken by  the number of helium atoms in the first solvation shell to reach an integer value $n_1(t)$ showed a linear dependence 
\begin{equation}
n_1(t) = At
\end{equation} 
 for the Ak$^+$He$_{n_1}$ complexes up to $n_1$ = 5, in agreement with the experimental result, pointing to a Poissonian process.

\begin{figure}[t]
\includegraphics[angle=270,width=1.0\linewidth,clip=true]{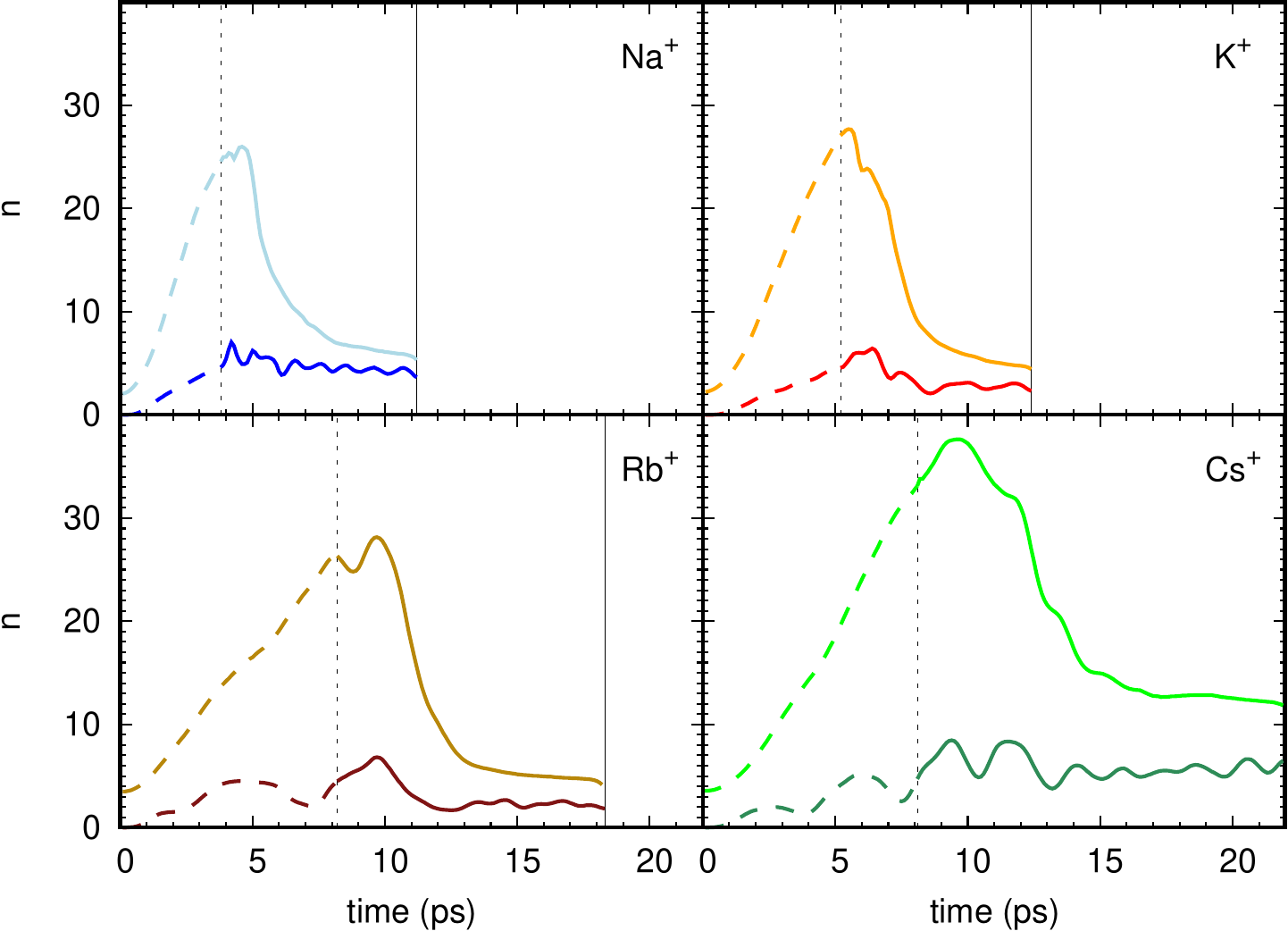}
\caption{\label{fig:n1n2-t-Ak-pump-probe}
Time evolution of $n_1$ and $n_1+n_2$, the number of helium atoms in the first and in the first and second solvation shells of Ak$^+$, during the pump (dashed lines) and the probe (solid lines) dynamics.
Darker, lower line: $n_1(t)$; lighter, higher line: $n_1(t)+n_2(t)$.
The pump-probe delay $\Delta t$ is indicated as a vertical dashed line: it corresponds to $n_1=5$ in the pump step and triggers the probe (ionization of Xe).
The solid line indicates the end of the simulation (when the second solvation shell reaches the absorption region in the simulation box). }
\end{figure}

Since we are now using a simulation grid with a smaller space step (0.2~\AA\ instead of 0.3~\AA), which makes it more accurate, especially in the case of the strongest Ak$^+$-He attractions (Na$^+$, K$^+$), we have redone the fit of the number of He atoms in the first solvation shell with a linear function of time.
The time evolution of the number of He atoms in the first ($n_1$) or second solvation structure ($n_2$) is plotted in  Fig.~\ref{fig:n1n2-t-Ak-pump-probe}, with the part to the left of the vertical dashed black line corresponding to the solvation (pump) step.
The linear behavior of $n_1(t)$ is clear for Na$^+$ and K$^+$, while some oscillations around it can be seen  for the heavier Rb$^+$ and Cs$^+$ ions.
This could be due to their larger size, allowing several He atoms to attempt binding at the same time. 
The new values of the slopes $A$ are collected in Table~\ref{Tab:R1n1dt}.
There are two different values of $A$ for the Rb$^+$ and Cs$^+$  ions: they are a result of the oscillations observed in the number of helium atoms in the first solvation shell.
Each value of the times at which the number of He atoms reached a given value of $n_1$ was determined in two different ways.
The first one was the first time at which there was $n_1$ atoms in the first solvation shell, while the second one was the time starting from which this number no longer fluctuated back to $n_1-1$.
The linear fit was conducted for each series.
In the case of Na$^+$ and K$^+$, there is no difference, at least up to $n_1=5$, whereas for the heavier Rb$^+$ and Cs$^+$ the differences are more important.
The values of $A$ show a decrease in the solvation rate from the most attractive (Na$^+$) to the least attractive (Cs$^+$) ion.

The only case that can be compared to experiment so far is that of Na$^+$.
The value of the binding rate of the first five He atoms found in our simulations is $1.33$~atom/ps, for an initial droplet size of $2000$~atoms.
This is about 20~\%\ lower than  the value of $1.65\pm 0.09$~atom/ps found by Albrechtsen et al.\cite{Albrechtsen2025} for $\langle N\rangle= 3500$.
This can be considered as a remarkable agreement since, as mentioned above, the experimental rate decreases with droplet size.
Note that our current result is somewhat larger than in our earlier study\cite{GarciaAlfonso2024} ($\left[0.79,0.74\right]$), thanks to the more accurate description of the helium density provided by the finer simulation grid.

\subsection{Probe step (Coulomb explosion) dynamics}\label{ssec:res-probe}

Given the cost of the simulations, we have chosen to simulate the Aarhus experiments\cite{Albrechtsen2023,Albrechtsen2025} for a value of the pump-probe time delay $\Delta t$ for which the number of helium atoms in the first solvation shell is equal to 5.
Beyond this value, the Na$^+$He$_n$ ion yields $Y_n(t)$ were influenced by the dissociation of higher Na$^+$He$_N$, $N>n$ complexes.
Consequently, the corresponding values of $\Delta t$ are equal to $3.8$, $5.2$, $8.2$, and $8.1$~ps for Na$^+$, K$^+$, Rb$^+$, and Cs$^+$, respectively.
These values are collected in Table~\ref{Tab:R1n1dt}.

As stated above, the probe step is simulated by assuming instantaneous ionization of the Xe atom, which triggers Coulomb repulsion between the ions and the ejection of Ak$^+$ from the drop.
Snapshots illustrating this dynamics in the case of Na$^+$  are shown in Figure~\ref{fig:probe-Nap-snapshots}.
The corresponding movie is included in the Supplementary Material as the second part of the overall movie including the pump step: 
the beginning of the probe step is marked by a change in color of the disk representing the Xe atom.
The same is true for the movies corresponding to the other alkalis (K$^+$, Rb$^+$, and Cs$^+$).

The 2-dimension density cuts on the top part of the figure help visualize the large amount of energy dissipated by the ionic complexes into the droplet, both as density waves traveling through the droplet and as surface excitations.
As in the case of the solvation (pump step)  dynamics, the lower part of the figure shows the helium density integrated in a sphere of radius $r$ around the Ak$^+$ ion, with the results for the first and second solvation shell radii, $r^e_1$ and $r^e_2$, highlighted as blue dashed lines.

Two conclusions can be drawn from these plots.
\\
- First, when $n_1=5$, \textit{i.e.} when the probe dynamics is triggered, the first solvation shell is not yet stabilized.
This could be expected since it can in principle hold up to $14$ atoms (Table~\ref{Tab:R1n1dt}).
In addition to its  incompleteness,  the solvation structure is very hot, as indicated by the strong density oscillations in the region of the first and second solvation shells, smearing out any possible distinction between them.
\\
- Second, the number of helium atoms inside the first solvation shell keeps evolving during the probe step.
This is due to three effects:
(i) It takes some time for the ionic complex to get out of  the droplet and  completely separated from it: 
The Ak$^+$ ion was moving towards the droplet center during the solvation (pump) step, and it velocity does not change orientation instantly; and the ion has to travel back out of the droplet during a certain time before getting out, which makes it possible for a few  more $^4$He atoms to enter the solvation shell.
(ii) The strong acceleration due to Xe$^+$--Ak$^+$ Coulomb repulsion makes the ionic complex collide with the surrounding He atoms, and presumably also pushes Ak$^+$ against the part of the solvation shell which is on its way outward while the other part is pulled toward the ion: this increases the internal energy of the solvation complex around the ion.
(iii) Given the high internal energy of the first solvation shell, evidenced by the multiple peaks of the integrated density (and by the energy analysis, see Section~\ref{sec:energy}), it can cool down by dissociating He atoms: this can occur inside as well as outside the droplet on the way to the detector.

Let us examine these points in more details.
The fact that the ion has penetrated deeply inside the droplet at the time where probe begins is evidenced by examining  the distance between Ak$^+$ and the overall center of mass as a function of time, plotted in Fig.~\ref{fig:Akp-Xep-HeCOMdist-t-all-Ak-pump-probe} during the pump and the probe steps (separated by a vertical bar in the figure).
All the Ak$^+$ ions have clearly penetrated a significant distance inside the droplet by the time the first solvation shell reaches a number of He atoms equal to 5.
This can be quantified from the values of the Ak$^+$--Xe distances collected in Table~\ref{Tab:R1n1dt} at the beginning ($d_0$) and the end [$d(\Delta t)$] of the pump stage.
For instance,  Na$^+$ has penetrated inside the droplet by 8.4~\AA ($d_0=31.9$~\AA, $d(\Delta t)=23.5$~\AA). 
This value decreases with increasing atomic number, being 7.2~\AA\ for K$^+$, 6.5~\AA\ for Rb$^+$, and 5.6~\AA\ for Cs$^+$.
It takes several picoseconds for the ions to go back to the position of the neutral atom at the beginning of the pump stage: 2.3~ps for Na$^+$, 2.6~ps for K$^+$, 3.5~ps for Rb$^+$, and 4.7~ps for Cs$^+$.
The time taken by the vertical component of the ion velocity to change sign is small in comparison, but can still be seen in Fig.~\ref{fig:Akp-Xep-HeCOMdist-t-all-Ak-pump-probe} as an offset between the start of the probe and the time at which the ion distance to the droplet center of mass (COM) starts increasing again.
Note that the ionic complex is still in contact with the droplet for another few picoseconds after that, as can be checked in the snapshot figure of the process, Fig.~\ref{fig:probe-Nap-snapshots} or from the movies in the Supplementary Material.

The variation of the number of He atoms attached to the ion during the probe step can be seen in Fig.~\ref{fig:n1n2-t-Ak-pump-probe}  introduced earlier, which represents the time evolution of the number of He atoms in the first ($n_1$) or second solvation structure ($n_2$).
During the probe step (Xe atom ionization, marked by a vertical dashed line in Fig.~\ref{fig:n1n2-t-Ak-pump-probe}), the number of He atoms in the first or second solvation structure does not instantly freeze, nor does it start decreasing as would be the case if Ak$^+$He$_n$ would start dissociating He atoms in order to dissipate its high internal energy. 
For instance, in the case of Na$^+$, the number of He atoms after the probe pulse ($t=\Delta t=3.8$~ps) grows from 5 to 7 during the first 0.4~ps, then oscillates while slowly decreasing until the end of the simulation down to 4 atoms.
The first shell seems to be more relevant at the beginning of the probe, since as can be seen in Fig.~\ref{fig:probe-Nap-snapshots} there is no clear separation between the second solvation shell and the rest of the droplet.
However, starting from the snapshot at 8.3~ps ($\Delta t+4.5$~ps), the whole solvation structure is included within this second shell and well separated from the rest of the droplet.
The total number of atoms in the second solvation structure is then equal to 7, and it decreases down to 5 at the end of the simulation and probably less if the simulation could be continued.

The reason for the increasing number of bound atoms  after the probe pulse stems from the observations made about Fig.~\ref{fig:Akp-Xep-HeCOMdist-t-all-Ak-pump-probe}.
Even though Ak$^+$ is strongly accelerated upon Xe atom ionization, it still takes a few picoseconds for it to  get out of the droplet and a little more to get separated from the droplet. 
Hence the outgoing ionic complex Ak$^+$He$_n$ keeps gaining a few He atoms during an additional few picoseconds.
After that, there is a strong decrease in the total number of He atoms accompanying the ion (which is well represented asymptotically by $n_2(t)$).

As can be realized by looking at the snapshots in Fig.~\ref{fig:probe-Nap-snapshots} for Na$^+$, or by viewing the corresponding movie and the ones for the other alkalis  in the Supplementary Material, the droplet is significantly distorted by the departure of the Ak$^+$ ion because of the strong Ak$^+$-He interaction, but also because of He-He interaction and correlation effects.
A "plume" of helium density follows the exiting Ak$^+$He$_n$ complex.
It corresponds  to some helium atoms from the droplet that are attracted by the charge, and to some dissociating from the solvation complex, both ending up  falling back on the droplet.

\section{Energy relaxation rate}\label{sec:energy}

In order to determine the energy $E_\text{dissip}$ dissipated during the solvation process from the local region around the ion to the rest of the droplet, Albrechtsen \textit{et al.}\cite{Albrechtsen2025} have used a model to fit the time-dependent size distributions of the detected Na$^+$He$_n$ ions.
The model made use of accurate theoretical values for the evaporation energies of the Na$^+$He$_n$ ions obtained by a path integral Monte Carlo method and a new potential energy surface.
It allowed to bracket the mean energy $\langle E_{\text{dissip}}\rangle (\Delta t)$ dissipated by the ionic complex during the solvation stage from the experimental distribution $P_\text{exp}(n;\Delta t)$ of Na$^+$He$_n$ ions at a pump-probe delay of $\Delta t$.

It was found that for the first 5~ps of the solvation process, the mean dissipated energy could be fitted with Newton's law of cooling, which assumes that the rate of energy transfer from the hot solvation structure to the surrounding helium droplet  is proportional to its internal energy:
\begin{equation}\label{eq:Newton-law-exp}
E^\text{Newt}_\textit{dissip} (t) = E_\text{dissip}(\infty) \left[ 1 - \exp\left(-\frac{t+\Delta t_\text{dissip}}{\tau_\text{dissip}}\right)\right]
\end{equation}
where $E_\text{dissp}(\infty)$ is the energy dissipated asymptotically, $\tau_\text{dissip}$ is the time constant for the energy dissipation, and $\Delta t_\text{dissip}$ is a time offset.\cite{Albrechtsen2025} 

Here we can determine directly the amount of energy $E_\text{solv.struct.}(t)$ contained in the solvation structure around the ion.
Comparing with Eq.~(6) in Ref.~\onlinecite{Albrechtsen2025}:
\begin{equation}\label{eq:Etotexp}
E_\text{init} (\text{Na}^+) = E_\text{dissip} + E_\text{bind}(N) + E_\text{int}(N),
\end{equation}
where $E_\text{init} (\text{Na}^+)$ is the total energy of the Na$^+$He$_{N_D}$ ion-drop system immediately after the sodium atom is ionized (at $t=0$), $E_\text{bind}(N)$ and $E_\text{int}(N)$ are the binding energy and the internal energy, respectively, of the Na$^+$He$_N$ complex at time $t$,
$E_\text{solv.struct.}(t)$  corresponds to 
\begin{equation}
E_\text{solv.struct.}(t) = E_\text{bind}(N) + E_\text{int}(N) = E_\text{init} (\text{Na}^+) - E_\text{dissip},
\end{equation}
with $N$ equal to the number of He atoms included in the ionic complex at time $t$.
Hence, $E_\text{init} (\text{Na}^+)$ being a constant, $E_\text{solv.struct.}$ should vary like $- E_\text{dissip}$.

We have  calculated the total energy  $E^{(1)}_\text{solv.struct.}$ or $E^{(2)}_\text{solv.struct.}$  of the solvation structure by integrating $\langle \Psi(t) \vert \mathcal{H} \vert \Psi(t) \rangle$ in a sphere around Na$^+$ with radius $r_1^e$ or $r_2^e$, respectively, $\Psi(t)$ being the time-dependent pseudo wave function described in Section~\ref{ssec:meth-dyn} for the solvation (pump) dynamics of Na$^+$He$_{2000}$ and $\mathcal{H}$ the corresponding hamiltonian.
For the purpose of this analysis, the solvation dynamics was extended beyond $t=\Delta t= 3.8$~ps at which probe starts.

\begin{figure}[t]
\includegraphics[angle=270,width=1.0\linewidth,clip=true]{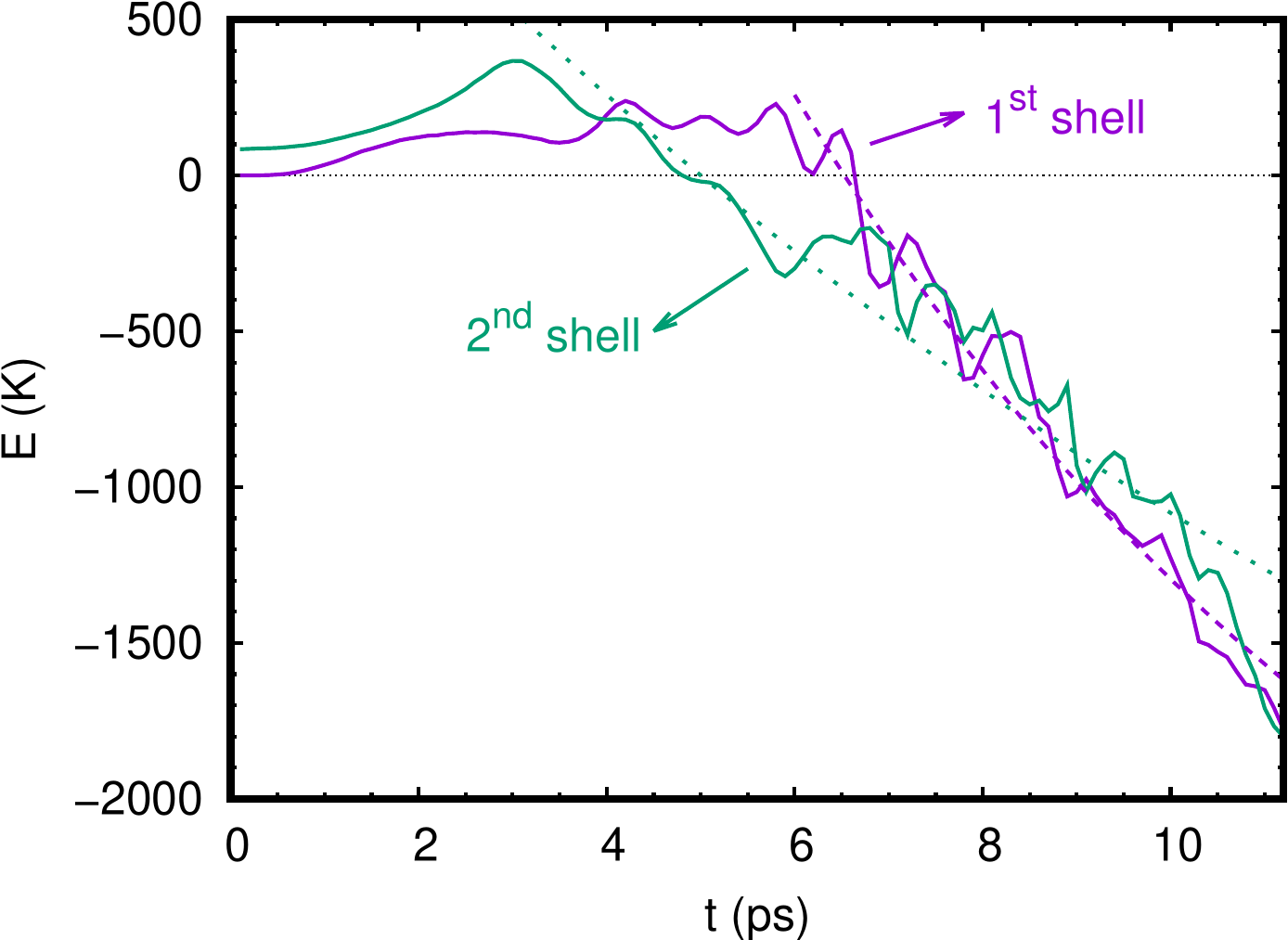}
\caption{\label{fig:fit-engy-Nap-FSS-SSS-pump}
Total energy  $E^{(n)}_\text{solv.struct.}(t)$ contained in the first  ($n=1$, purple curve) and second ($n=2$, green curve) solvation structures around Na$^+$ during the pump (solvation) stage of Na$^+$ in a Xe-containing $^4$He$_{2000}$ droplet. 
Also shown are the results of fitting these time evolutions by an exponential decay (Newton law, see text), after a given time delay:
dashed, purple line for the first and dotted, green line for the second solvation structure, respectively.
}
\end{figure}

The results of the time evolution of the energy contained in the solvation structure are shown in Fig.~\ref{fig:fit-engy-Nap-FSS-SSS-pump}. 
As can be seen in that figure, $E^{(n)}_\text{solv.struct.}(t)$ behaves similarly for $n=1$ or 2:
this means that our results are robust with respect to the definition of the solvation structure. 

The beginning of the time evolution of $E^{(n)}_\text{solv.struct.}(t)$ came as a surprise: the energy is positive and increasing, for a significant amount of time. 
Indeed, it takes about 5 ps for $E^{(2)}_\text{solv.struct.}$ and 6.5 ps for $E^{(1)}_\text{solv.struct.}$ to become negative: 
 the solvation structure is not stable by itself. 
This is due to the large difference between neutral Na-He and ionic Na$^+$-He interactions. 
The binding distances and energies are very different, see Table~\ref{Tab:pots}. 
Hence upon  Na atom ionization, the surrounding helium atoms are strongly attracted to the charge and collide with the  ion and with each other, and rebound and collide with the outer helium atoms, which corresponds to a very high internal energy in the solvation structure being formed. 
The very strong helium-charge attraction of the He atoms in the solvation complex is overcome by their very high kinetic energy.  
This is the reason for the multiple peaks observed in the integrated density around the ion at the early stages of the dynamics (Fig.~\ref{fig:pump-Nap-snapshots} and corresponding movie in the Supplementary Material). 
This means that the stability of the structure and its drive towards the droplet center is actually provided by the rest of the droplet, thanks to the long range charge-induced dipole attraction. 
At longer  times, $E^{(n)}_\text{solv.struct.}(t)$  becomes negative: enough energy has been dissipated by the solvation structure in formation that it has become stable by itself, and it keeps relaxing towards its equilibrium structure.

In order to compare with experimental results, we have fitted this energy relaxation with Newton's law of cooling: 
\begin{equation}\label{eq:Newton}
E^{\text{Newton}} = E_\infty  \left\lbrack 1-\exp\left( - \frac{t-t_0}{\tau } \right) \right\rbrack.
\end{equation}
In this equation, $E_\infty$ is the asymptotic energy, \textit{i.e.}, the energy of the complete (first or second) solvation structure at equilibrium, which was determined separately. 
We have determined $t_0$ and $\tau$ using a least-square fit starting from the time at which the energy started to decrease.
The results of the fits are  illustrated in Fig.~\ref{fig:fit-engy-Nap-FSS-SSS-pump} and the resulting parameters are collected in Table~\ref{Tab:Newton-fit}.


 \begin{table}
\begin{center}
\setlength{\tabcolsep}{6pt} 
\begin{tabular}{cccccc}
\hline
\hline
 & $t_0$ & $\tau$  & $E_\infty$ &  interval        & rms  \\
 &  (ps)   &   (ps)    &  (K)           &  (ps)             &  (K)   \\
\hline
$E_\text{solv.struct.}^{(1)}$  &$6.53\pm0.06$  & $7.3\pm0.2$ & $-3424$ &  [6.0,11.2,] & 112  \\
$E_\text{solv.struct.}^{(2)}$  & $5.0\pm 0.1$  & $16.5\pm 0.6$ & $-4144$ & [3.0,11.2] & 162  \\
\hline
\end{tabular}
\end{center}
\caption{
Parameters and results from fitting the energy integrated in the first or second solvation structure with Newton's law [Eq.~\ref{eq:Newton}].
\label{Tab:Newton-fit}
}
\end{table}

As can be seen from  Table~\ref{Tab:Newton-fit}, 
the larger the limits used to define the solvation structure, the earlier the total energy becomes negative.
This is accompanied by a slower relaxation rate.
The difference can be considered as a error margin in the determination of these parameters.

Comparing with the results presented by Albrechtsen \textit{et al.}\cite{Albrechtsen2025} from the analysis of their experimental results, it is quite satisfying that we indeed find a behavior confirming Newton's law for energy dissipation.
Unfortunately, the parameter values of  Albrechtsen \textit{et al.}\cite{Albrechtsen2025}  are rather different:
for the size distribution $\langle N\rangle=3600$, which is the one closest to our simulated size of 2000,
they obtained: 
$\tau=2.6\pm 0.4$~ps; 
$t_0$ [$=-\Delta t_{dissip}$ in Eq.~(\ref{eq:Newton-law-exp})] $=0.23\pm0.06$~ps. 
The parameters $\tau$ and $t_0$  can be compared directly.
Our value of $t_0$ (5.0~ps or 6.5~ps) is much larger than the one deduced from the experimental model,
and the relaxation rate is  slower in our simulation ($\tau = 7.3$ or $16.5$~ps). 

The difference between experimental and simulated results is puzzling.
It may be due to the difficulty of defining the spatial limits of the solvation shell in the simulation.
Indeed, the radius of the equilibrium first solvation shell may appear to be too small, since helium atoms (density) keep coming in and out because of the high internal energy, which is presumably responsible for the oscillations observed for $E(t)$ in the top plot of  Fig.~\ref{fig:fit-engy-Nap-FSS-SSS-pump}.
However, including a wider spatial region by using the second solvation structure 
makes Newton's behavior start earlier, but the difference with the other data (total energy dissipated and energy dissipation rate)  even larger.
It could of course be due to shortcomings of the He-TDDFT description of the dynamics, although it has been successful in describing a number of other dynamics processes. 
The differences could also originate from some of the hypotheses used in the experimental model.
For instance, as discussed in Section~\ref{ssec:res-probe}, our simulations of the probe step show that it takes a significant amount of time after the Xe atom ionization for the ionic complex to get out of the droplet and even more to separate completely from it.
During that time, the ionic complex keeps exchanging energy and He atoms with the droplet, which could not be taken into account in the experimental model.



\section{Conclusion}\label{sec:conclusion}
We have conducted a theoretical  study of the pump and probe steps of the experiment by Albrechtsen \textit{et al.}\cite{Albrechtsen2023,Albrechtsen2025} on the solvation of an alkali ion, Na$^+$, K$^+$, Rb$^+$ and Cs$^+$, in a $^4$He$_{2000}$ droplet using the $^4$He-TDDFT approach.
Our results for the solvation dynamics (pump step) are similar to the ones published earlier\cite{GarciaAlfonso2024}, although no Xe atom was present.
Note that the  accuracy is improved in this work, especially for the lighter alkalis Na$^+$ and K$^+$, thanks to a denser simulation grid.

The Ak$^+$ ion is found to penetrate by several Angstroms inside the droplet as more and more He atoms bind to it.
The Ak$^+$He$_n$ complex being formed is very hot, as testified by a highly structured density profile around the ion.
This is a consequence of the large energy difference between the initial conditions, the Ak$^+$ ion created at the droplet surface, and the equilibrium configuration of the solvated ion at the center of the droplet.
It reflects the large difference between the neutral Ak-He and ionic Ak$^+$-He interactions. 

Our results for the binding rate of He atoms to the Na$^+$ ion compare very well with the new experimental  ones.\cite{Albrechtsen2025} 
They confirm that the binding of the first five atoms follow a Poisson model for Na$^+$ and K$^+$, and in a lesser measure also for Rb$^+$ and Cs$^+$, for which some oscillations around the linear behavior can be seen, presumably due to their bigger size allowing several He atoms to start binding at the same time.

The simulation of the probe step, triggered by ionizing the central Xe atom, was conducted for a pump-probe delay such that the solvation complex had reached five He atoms.
Two important conclusions could be drawn.
First, the Ak$^+$He$_5$ solvation complex was far from being stabilized at the time of the probe: it was very hot, as witnessed by the highly structured density profile around the ion.
Second, the number of helium atoms inside the solvation structure kept evolving during the probe step.
This was due to the time taken by the Ak$^+$ ion to turn around and travel back the distance it had penetrated inside the droplet, during which it had time to gain more He atoms, due to collisions with the surrounding helium on its way out, and to cooling of the high internal energy by He atom evaporation once the ionic complex was separated from the rest of the droplet.

The analysis of the energy content of the solvation shell conducted for Na$^+$ solvation has revealed the surprising result that it is not stable by itself during the first few picoseconds, but only stabilized by the surrounding helium solvent.
After that, energy dissipation follows Newton's law as concluded by experiment\cite{Albrechtsen2025}, although with a slower rate.

\section{Supplementary Material}

Movies illustrating the pump-probe process for Na$^+$ (Multimedia available online), K$^+$ (Multimedia available online), Rb$^+$ (Multimedia available online) and Cs$^+$ (Multimedia available online) are included in this Supplementary Material.
They were obtained from the liquid $^4$He time-dependent density functional theory ($^4$He-TDDFT) simulations at zero temperature presented in the main text.
The alkali atom Ak$\equiv$ Na, K, Rb or Cs,  initially resides at the surface of a 2000-$^4$He atom droplet containing a xenon atom at its center.
Ak is suddenly ionized at time $t=0$ by the pump pulse.
After a given time delay $\Delta t$ at which the Ak$^+$   ion first solvation shell contains five helium atoms, the central Xe atom is ionized by the probe pulse, 
which triggers Coulomb repulsion between the two ions and ejection of the ionic alkali solvation complex.

The movies show 2 panels (see figures in the following pages).
The left-hand panel describes the time-evolution of a 2-dimension cut through the droplet density in the $(x,z)$ plane with the Xe atom initially at the droplet center and the Ak atom initially at the droplet surface.
The density scale is given on the right in units of $\rho_0=0.0218$~atom/\AA$^{-3}$, the bulk superfluid helium density at temperature $T=0$~K.
Ionization of the Xe atom is denoted by the change of color of this central atom from grey to blue.
The right-hand panel shows the spherically-integrated density around  the alkali ion as a function of the distance to the ion, with the corresponding number of He atoms included  (dashed red line referred to the right-hand scale);  the results for the first and second solvation structures are stressed as dashed blue lines.

 
\begin{figure*}
\includegraphics[width=0.98\linewidth,angle=0,clip=true]{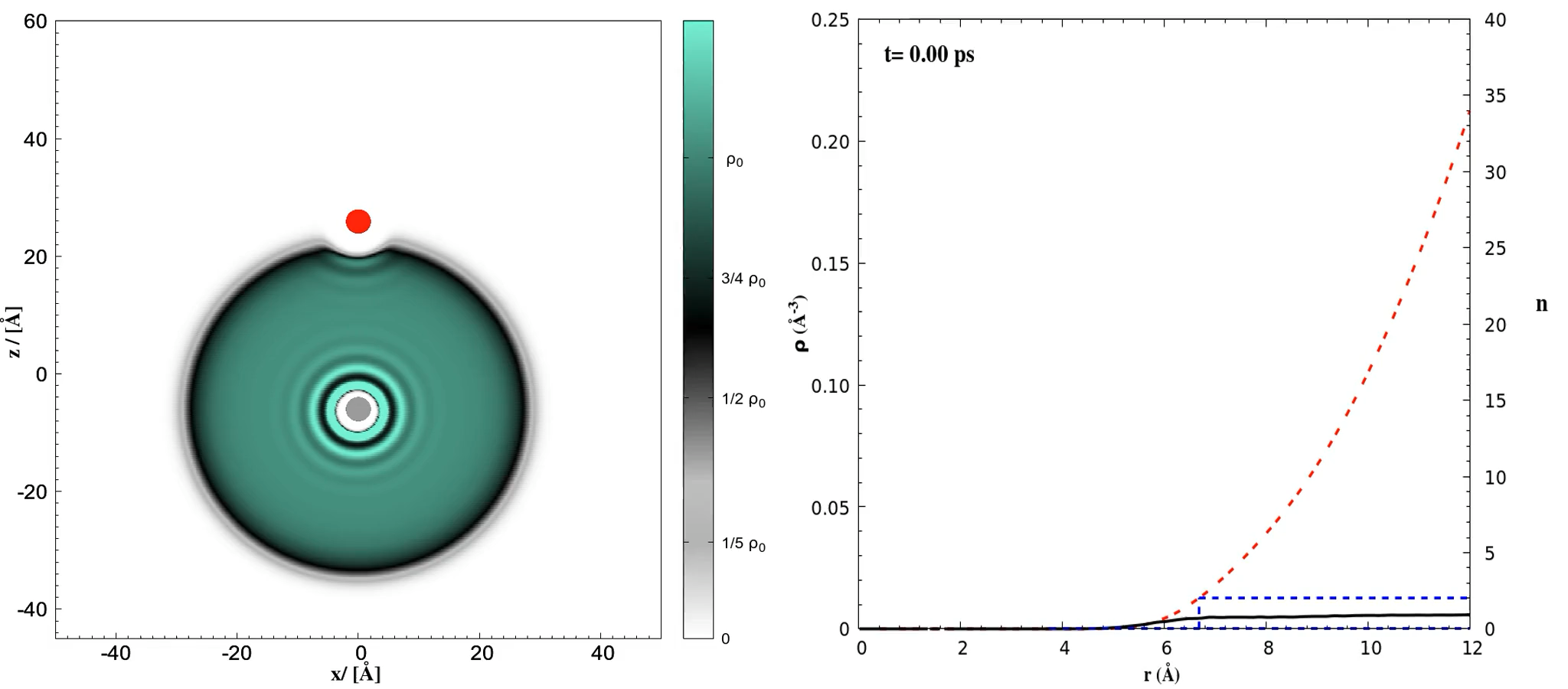}
\caption{\label{fig:Nap-XeHe2000-pump-probe}
First part: Solvation of Na$^+$@(Xe@$^4$He$_{2000}$ (pump step);
Second part (marked by the change of color of the central Xe atom from grey to blue): At $t=\Delta t=3.8$~ps, \textit{i.e.} when the Na$^+$ ion first solvation shell contains five $^4$He atoms, the Xe atom is ionized (probe step) triggering Coulomb repulsion between the ions.
See explanation of the two panels in the Description section.
(Multimedia available online) }
\end{figure*}


 
\begin{figure*}
\includegraphics[width=0.98\linewidth,angle=0,clip=true]{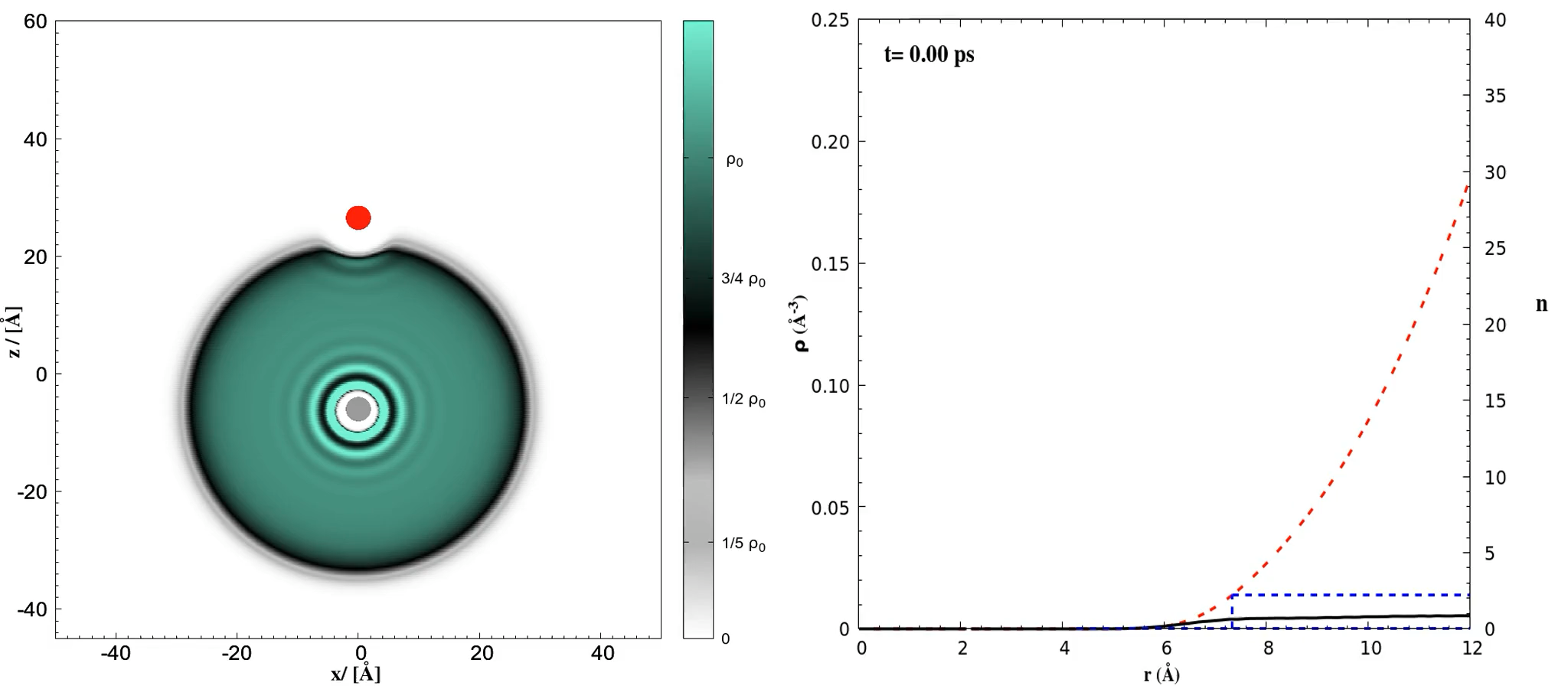}
\caption{\label{fig:Kp-XeHe2000-pump-probe}
First part: Solvation of K$^+$@(Xe@$^4$He$_{2000}$ (pump step);
Second part (marked by the change of color of the central Xe atom from grey to blue): At $t=\Delta t=5.2$~ps, \textit{i.e.} when the K$^+$ ion first solvation shell contains five $^4$He atoms, the Xe atom is ionized (probe step) triggering Coulomb repulsion between the ions.
See explanation of the two panels in the Description section.
(Multimedia available online) }
\end{figure*}


 
\begin{figure*}
\includegraphics[width=0.98\linewidth,angle=0,clip=true]{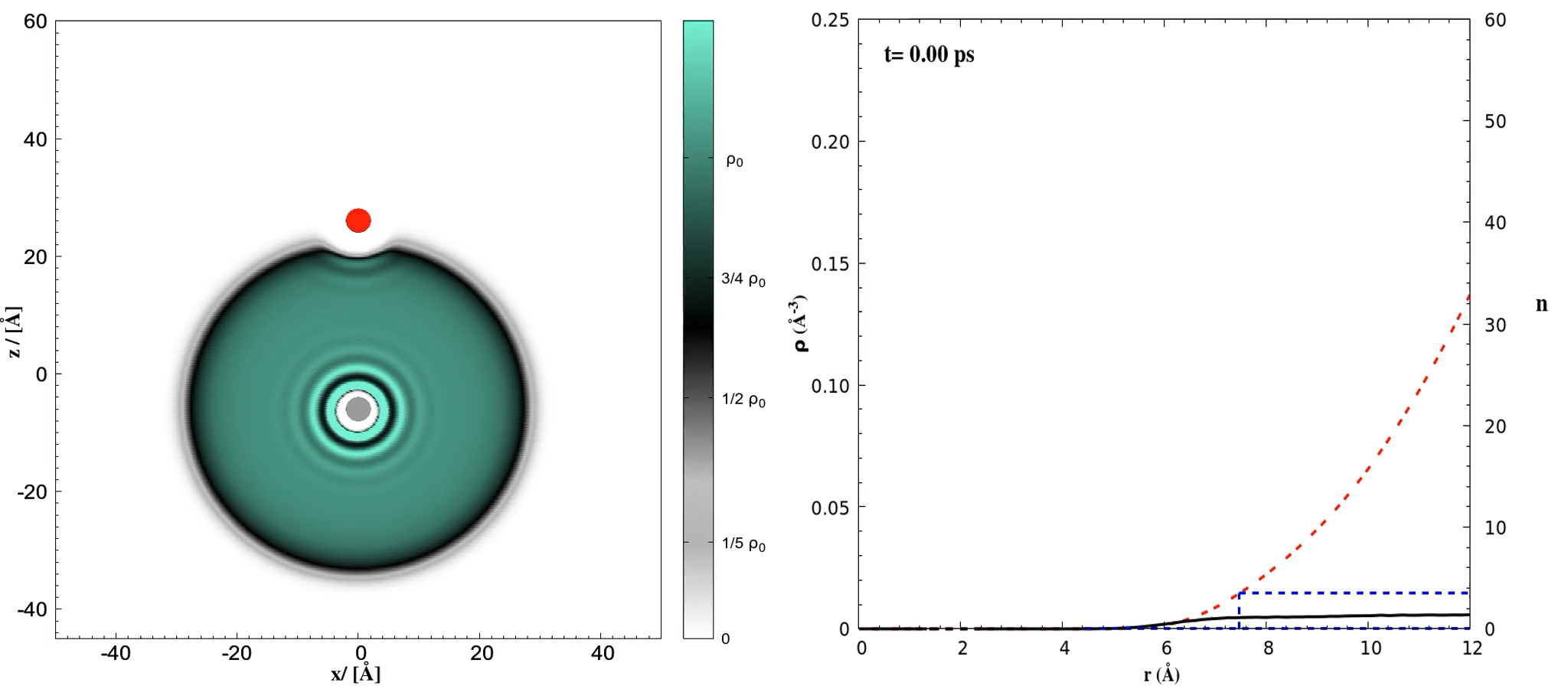}
\caption{\label{fig:Rbp-XeHe2000-pump-probe}
First part: Solvation of Rb$^+$@(Xe@$^4$He$_{2000}$ (pump step);
Second part (marked by the change of color of the central Xe atom from grey to blue): At $t=\Delta t=8.2$~ps, \textit{i.e.} when the Rb$^+$ ion first solvation shell contains five $^4$He atoms, the Xe atom is ionized (probe step) triggering Coulomb repulsion between the ions.
See explanation of the two panels in the Description section.
(Multimedia available online) }
\end{figure*}


 
\begin{figure*}
\includegraphics[width=0.98\linewidth,angle=0,clip=true]{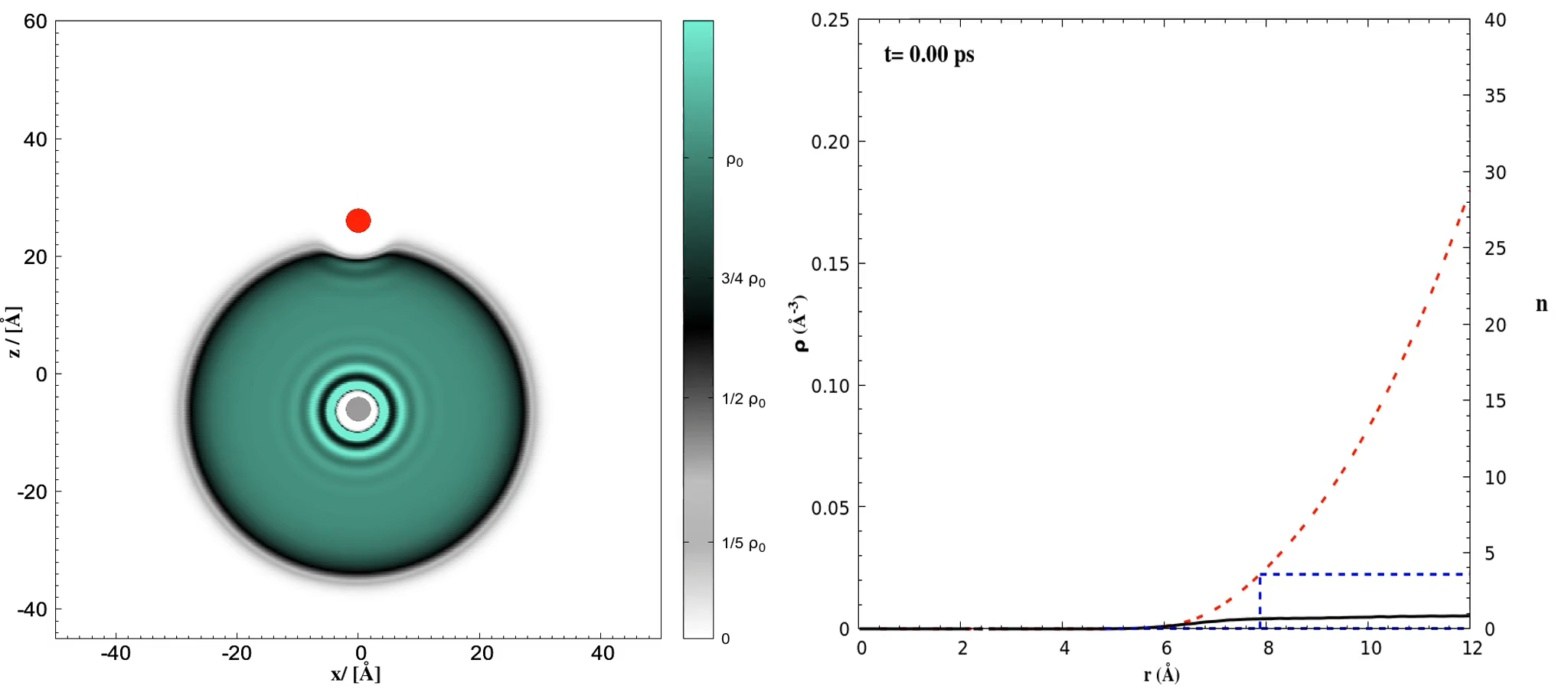}
\caption{\label{fig:Csp-XeHe2000-pump-probe}
First part: Solvation of Cs$^+$@(Xe@$^4$He$_{2000}$ (pump step);
Second part (marked by the change of color of the central Xe atom from grey to blue): At $t=\Delta t=8.1$~ps, \textit{i.e.} when the Cs$^+$ ion first solvation shell contains five $^4$He atoms, the Xe atom is ionized (probe step) triggering Coulomb repulsion between the ions.
See explanation of the two panels in the Description section.
(Multimedia available online) }
\end{figure*}

 \begin{acknowledgments}
 A computer grant from CALMIP high performance computer center (grant P1039) is gratefully acknowledged.
  This work has been funded by Grant PID2023-147475NB-I00 funded by MICIU/AEI/10.13039/501100011033 
and benefited from COST Action CA21101 ``Confined molecular systems: form a new generation of materials to the stars'' (COSY) 
supported by COST (European Cooperation in Science and Technology).
\end{acknowledgments}


\section*{AUTHOR DECLARATIONS}   
\subsection*{Conflict of Interest}
The authors have no conflicts to disclose.


\subsection*{DATA AVAILABILITY}
The data that support the findings of this study are available
from the corresponding author upon reasonable request


\clearpage


 \section*{REFERENCES}

\bibliographystyle{apsrev4-1}
\bibliography{../../../../../biblio/HeN/Hecluster_complete2.bib}

\end{document}